\documentclass[10pt, conference]{IEEEtran}
\IEEEoverridecommandlockouts
\usepackage{cite}
\usepackage{amsmath,amssymb,amsfonts}
\usepackage{algorithmic}
\usepackage{graphicx}
\usepackage{textcomp}
\usepackage{xcolor}
\usepackage{hyperref}
\usepackage{cleveref}

\def\BibTeX{{\rm B\kern-.05em{\sc i\kern-.025em b}\kern-.08em
    T\kern-.1667em\lower.7ex\hbox{E}\kern-.125emX}}
\begin{document}
\newcommand{\etal}{\textit{et al.}}

\title{An architecture for enabling A/B experiments in automotive embedded software
\thanks{}
}

\author{\IEEEauthorblockN{Yuchu Liu}
\IEEEauthorblockA{
\textit{Volvo Cars}\\
Gothenburg, Sweden \\
yuchu.liu@volvocars.com}
\and
\IEEEauthorblockN{Jan Bosch}
\IEEEauthorblockA{\textit{Computer Science and Engineering} \\
\textit{Chalmers University of Technology}\\
Gothenburg, Sweden \\
jan.bosch@chalmers.se}
\and
\IEEEauthorblockN{Helena Holmstr\"om Olsson
\IEEEauthorblockA{\textit{Computer Science and Media Technology} \\
\textit{Malm\"o University}\\
Malm\"o, Sweden \\
helena.holmstrom.olsson@mau.se}
\and
\IEEEauthorblockN{Jonn Lantz}
\IEEEauthorblockA{
\textit{Volvo Cars}\\
Gothenburg, Sweden \\
jonn.lantz@volvocars.com}
}}

\maketitle

\begin{abstract}
A/B experimentation is a known technique for data-driven product development and has demonstrated its value in web-facing businesses.
With the digitalisation of the automotive industry, the focus in the industry is shifting towards software. 
For automotive embedded software to continuously improve, A/B experimentation is considered an important technique.
However, the adoption of such a technique is not without challenge.
In this paper, we present an architecture to enable A/B testing in automotive embedded software. 
The design addresses challenges that are unique to the automotive industry in a systematic fashion.
%We share the critical components of the design and describe the system characteristics.
Going from hypothesis to practice, our architecture was also applied in practice for running online experiments on a considerable scale.
Furthermore, a case study approach was used to compare our proposal with state-of-practice in the automotive industry.
We found our architecture design to be relevant and applicable in the efforts of adopting continuous A/B experiments in automotive embedded software.
\end{abstract}


\section{Introduction \label{intro}}

A/B experimentation or A/B testing is a method for evaluating software changes in a quantifiable manner.
Continuous A/B testing is an important method in understanding and delivering measurable customer value.
Many web-facing companies have demonstrated success from A/B experiments, such as Booking.com \cite{Fabijan2018}, Google \cite{google2010} and Microsoft \cite{Kohavi2013, Gupta2018, Li2019}, just to list a few.
With the digitalisation of the automotive industry, software is becoming a main differentiator of products \cite{Mattos2018}. 
A/B testing is an effective tool to evaluate software and support organisations in making data-driven decisions \cite{Fabijan2017a}. 
%More and more automotive functions are becoming difficult to validate in a lab-like environment \cite{Koopman2016}. 
However, the adoption of continuous A/B experiments in automotive embedded software is not without challenges.

Embedded software has hardware constraints. Such constraints could manifest as limitations to computational power \cite{Giaimo2017}, long release cycles \cite{Mattos2018} and often dependency on suppliers \cite{Mattos2020}. 
Data collection and handling is also believed to be challenging in the automotive specific applications \cite{Giaimo2019, Mattos2020}. 
Although a fair number of publications point out the challenges in A/B experiment adoption \cite{Giaimo2017, Mattos2018, Giaimo2019, Mattos2020}, we identified a gap in the literature concerning architectural solutions to enable A/B experiments.
Furthermore, there is little to no reports on concluded or ongoing online A/B experiments in the automotive domain.
%We identified a gap in the literature on suitable A/B experiment architecture for automotive software.

In this paper, we present an architecture that enables A/B experiments in the automotive domain and aim to address the challenges that are unique to this industry.
We present a literature review of A/B experiment architecture in embedded and web-facing environments.
Moreover, we conducted a case study of the architecture applied at scale and to report the state-of-practice of A/B testing in automotive.
%Moreover, we studied a number of similar architectures on small to medium scales in three separate automotive companies.
Compared to the existing literature, the contribution of this paper is two-fold.
First, we present an architecture that enables A/B testing automotive software.
We reviewed the literature and did not find a similar architecture for A/B experiments.
Secondly, we apply this architecture in practice, in fleets of considerable scale. 
We present the case study and state-of-practice of two other automotive companies.

The rest of this paper is organised as following. 
In \cref{background}, we introduce the unique constraints in automotive industry for A/B testing. 
In \cref{method}, we present our research method.
We summarise the existing A/B experiment frameworks and architecture in \cref{otherworks}.
In \cref{design_main}, we present our architecture design along with the case studies. 
Discussions and conclusion are presented in \cref{discussion} and \cref{conclusion}.

\section{Background and constraints \label{background}}

In this section, we introduce the background on A/B testing and list the constraints of adopting the method in automotive embedded software.

\subsection{Background}

A/B testing is a type of continuous experimentation where users or systems are split into subgroups and issued with different variants of the same software.
By studying the response from each cohorts, A/B experiments can guide product development in an effective manner \cite{Kohavi2013, google2010, Fabijan2018}.
Typically, eligible users are split into two groups, the A version (control) and the B version (treatment). 
%The user group allocations can be done either in real-time \cite{Urban2016} or predetermined \cite{Kohavi2013}. 
For both user groups, their interactions with the functions are recorded and evaluated based on a set of carefully designed metrics reflecting business and/or customer values \cite{Dmitriev2017}.
%Invalid experiments which have a negative impact on user experience should be automatically eliminated \cite{Kohavi2013, google2010}.

Almost all well-established A/B testing frameworks are for web-facing businesses. 
Such frameworks or models cannot be applied directly in an embedded environment as they do not address specific challenges. 
These challenges come from many aspects, they can be technical, business, and organisational as demonstrated by Mattos \etal\ \cite{Mattos2018}. 
As embedded software often has dependency on hardware, fast software release becomes difficult to accomplish \cite{Giaimo2017, Giaimo2019, Mattos2020}. 
%While some concerns are expressed over safety and the legal frameworks \cite{Giaimo2017}, some concludes that A/B experiments do not necessarily have an impact on safety and quality, provided that the B variants released go through the same validation and verification processes \cite{Mattos2020}.
Although challenging to adopt, many advantages of continuous experiments that were proven in the web-facing businesses are also expected in the automotive industry \cite{Giaimo2019}. 
    
    \subsection{Constraints}
    
In addition to the challenges summarised by relevant literature \cite{Mattos2018, Giaimo2019, Mattos2020}, we list the specific constraints in automotive which motivate our architecture design. 
Automotive embedded software is distributed to hundreds of Electronics Control Modules (ECUs).
%which control sensors and actuators through low-level code.
These software are traditionally developed using the "V-model" where the OEMs deliver specifications and suppliers deliver implementations \cite{Forsberg1992}.
%The functional testing is not performed until the end of such development phase, with suppliers taking a large part of the development responsibility.
This model has exhibited its limitations.
%Due to the cost competitiveness of automotive business, software in this industry is highly cost-sensitive. 
%As suggested by Broy \etal\ \cite{Broy2007}, the traditional ECU unit-based cost model restricts versatility of software changes, as ECUs are often optimised in terms of computational power and memory space.
    
    \subsubsection{Release cycles and speed\label{background_speed}}

%Automotive software have grown from a few megabytes per vehicle to well over one gigabyte in just two decades.
Combining the strict standards with the growing complexity, the automotive software release process is rigid.
First, the development and release of automotive embedded software is usually strongly dependent on suppliers.
%whose time plan and cost can determine the release speed and cycle.
%This supplier-OEM relation can influence software that is internally developed when functions have dependency on one another. Such a challenge is often more organisational than technical.
Secondly, automotive companies have traditionally designed software release cycles based on their hardware release process \cite{Bosch2012}.
This process cannot handle rapid changes, as all integration and tests are planned at fixed periods.
%If software changes occur after a designated testing period, these changes would be scheduled to the next release slot thus hindering the release speed.
Moreover, the most commonly adopted automotive software architecture AUTOSAR \footnote{\href{https://www.autosar.org/}{https://www.autosar.org/}} lacks flexibility in partial updates \cite{Mattos2020}.
%Any minor change in the software requires a complete update of the entire ECU. 
If the new software is not backwards compatible, all ECUs in the vehicle need to be updated.
%In addition to the lack of rapid release processes, only updating specific ECU or several ECUs and ensuring backwards compatibility remains a challenge.
Last but not least, updating software which are governed by legislation might require renewal of certifications, which will add delays to the software release process.
%Whether the renewal is administrative or requires re-conducting certification tests witnessed by legal services, it 

    \subsubsection{Sample size and management\label{background_sample_size}}
Controlling boundary conditions is impossible for online experiments, as vehicles can be driven to everywhere and at anytime.
%As different treatments are applied to sample groups with many boundary conditions that are impossible to measure or quantify
Therefore, to conclude sufficient treatment effects, A/B experiments need be conducted on large and randomly selected sample groups.
This large group of users needs to be managed as online experiments require a flexible configuration of A/B or A/B/n groups.
%Users sometimes need to be stratified by certain properties of the vehicles such as model or engine types.
%to test variability of the response. 
%A study conducted by Jiang \etal\ \cite{Jiang2016} points out when A/B sample groups users are correlated or interacting, they will generate network or spill-over effects in the measurements. 
%An intuitive solution is to re-partition the sample groups and apply the same treatments. 
However, the sample groups are difficult to manipulate when the software needs to be updated through physical contact with the cars. 
Same challenge could be experienced when an A/B test is concluded, and the software needs to be inverted to the original version.

Managing sample groups longitudinally can be burdensome. 
Performance of some automotive functions depends on temporal factors and has seasonality effects, thus experiments need to be conducted longitudinally.
%Moreover, the duration of A/B experiments cannot be predetermined.
%as the time it will take to reach sufficient treatment effect is hard to predict prior to the experiments. 
%To avoid the risk of undetectable treatment effects, most A/B tests should not be scheduled to end but terminated only when a conclusion is reached. 
Therefore, the ability to orchestrate the A/B groups over time is beneficial.

    \subsubsection{Data infrastructure\label{background_data}}
To conclude a casual effect of the treatment, data collection for A/B experiments requires certain level of accuracy.
%In automotive applications, most experiments require time series data of trips instead of aggregated values per trip or snapshot representations of the trip. 
%Meratnia and By \cite{Meratnia2004} show that data collection from moving objects is expensive, specially when the vehicle is generating data at a high frequency. 
Storing such data locally in each vehicle is not feasible, as it becomes difficult to access and it will require a large memory on-board.
The success of an A/B experiment is largely relied on appropriate assumptions when designing an experiment and fast feedback when conducting one.
Sharing data within a large organisation can be problematic \cite{Fabijan2016}.
In order to maximise the data, all development teams need to have easy access to relevant data.
%Research from Fabijan \etal\ \cite{Fabijan2016} shows sharing of data within an organisation can be problematic despite having large amount of customer data collected. 
As a result, companies suffer from misrepresentation of customer values. 
%which are relevant to their functions and features. 
%The data to the development feedback loop often falls short, thus creating the "open-loop" problem as highlighted by Olsson and Bosch \cite{Olsson2014}. 

\subsubsection{Safety requirements and fallback\label{background_fallback}}

Automotive software has high safety requirements.
%A standardised method for risk classification is Automotive Safety Integrity Level (ASIL) \cite{ISO26262}.
%ASIL defines risk as a product of severity, controllability and event likelihood. To validate compliance of such risk classification, all the evaluations conducted need to be documented in the software itself. 
%Any functional changes will require new safety tests and analysis thus creating a long lead time. 
In an A/B test, all alternative versions can never obstruct such requirements which might affect road safety and/or legal compliance. 
The functional requirements need to be safeguarded while ensuring a continuous release of alternative versions seems impossible today.
Another practice to decrease hazards on the road is to have built-in fallback for safety critical functions.
For instance, one could install both the A and B alternatives on-board. Then the A alternative can be used as a fallback when it is thoroughly tested and validated.
%However, this practice might not be feasible due to the hardware limitations such as computation power and memory.

\section{Research method \label{method}}

In this paper, we combine a literature review with case studies. 
%The architecture we present in this paper is driven by experience in the automotive industry.
We studied several existing A/B experiment frameworks inside and outside of the industry through literature reviews, to compare our approach to existing frameworks.
%The insights helped us in summarising constraints and formulating requirements of A/B testing within the automotive domain.
Furthermore, to validate the architecture designed, we conducted case studies based on a series of ongoing efforts in A/B experiments from three separate automotive manufacturers.

We explore the following research question:

\begin{itemize}
  \item[] \textbf{RQ} How can we continuously experiment with automotive embedded software providing the challenges and limitations that are unique to this industry?
\end{itemize}

    \subsection{Literature review}
 
This literature review is done to understand existing A/B experiment frameworks within and outside of the automotive domain.
To identity and explore work that is relevant for the research question, we follow the methodology described by Kitchenham \cite{Kit04}.

    \subsubsection{Data collection}
    
We included the following terms in our search query: ("A/B testing" OR "A/B experiment" OR "online experiment" OR "bucket testing" OR "continuous experiment") AND ("software architecture") AND ("embedded software" or "automotive software").
Alternative terms are included as there is no standard terminology.
Keyword combination with "automotive software" yield no meaningful results, thus we expanded the search query to also include embedded software.
The databases included in our search process are IEEE Xplore, ScienceDirect, and Google Scholar, returning a total of 104 results excluding duplicates.
%We include both journal and conference papers in the search.
To ensure the results are relevant today, we limit the publications to the recent ten years.

    \subsubsection{Inclusion criteria}

Each paper resulted from the search process was reviewed by at least one of the authors. 
We examine the keywords, abstracts, and the body of the paper to identify A/B experiment frameworks and the applicable sector for said frameworks. 
We selected publications which focus on A/B experiment architecture and/or framework from embedded applications. 
We did not include publications discussing the benefits or challenges or feasibility of A/B testing.
%, nor did we include reports on conducted experiments or statistical methods for A/B experiments. 
This inclusion criteria resulted in a total of three papers. 
Since the technique is well established in web-facing applications, we included work on A/B testing framework in the web domain.
A total of 11 publications included in this review are \cite{google2010, Eklund2012, Bosch2012, Amatriain2013, Kohavi2013, Giaimo2017, Fagerholm2017, Fabijan2018, Gupta2018, Vasthimal2019, Li2019}.

    \subsection{Case study}

Following guidelines from Runeson and H\"{o}st \cite{Runeson2008}, we conducted two sets of case studies with three separate automotive companies. 
In study I, we examine the proposed architecture in practice on a cloud-based A/B experimentation in a vehicle fleet at scale. 
%We conducted a second case study through semi-structured interviews and meetings with two other automotive companies, to compare their current state-of-practice to our architecture design.
We study the architecture for A/B testing in a fleet from one of the three companies.
The software for case study I was developed in-house in company A.
%We followed the architecture to a large extent in the ongoing experiment, from software release, cloud/local parameter sets to cloud-based data collection.
As online experiments are not commonly applied in the industry, to the best of our knowledge, there is a lack of quantitative data to study from.
To understand the state-of-practice, we conducted semi-structured interviews with two more OEMs as case study II.
    
    \subsubsection{Case study attendees}
    
The three companies included in the case studies are large OEMs. 
%Their business includes design, development, and manufacturing of passenger and commercial vehicles. We address the three companies anonymously as A, B, and C.
In each company, we conducted interviews and workshops with at least five different employees from each company, working with varying aspects of software development. 
Their roles include software engineer, software architect, product owner, data engineer and data scientist.

    \subsubsection{Data collection}
    
One of the authors was actively involved in the experimentation design from ground up and supported the entire process.
We document the process through meeting notes and design specifications in the project. 
%The eligible users included in the fleet are corporate customers who use the vehicles as their primary family vehicles on a daily basis. 
%This A/B experiment is designed to evaluate a certain energy management function of electrified vehicles and is of a confidential nature.
%Therefore, in this paper, we will describe the architecture and setup instead of the actual A/B experiment and their results. 
The questions from case study II were specifically designed to understand the current state-or-practice of A/B experiments in an automotive setting.
We also aim to understand the potential of cloud-based A/B testings in each company.
During the interviews, we presented our architecture design to the attendees along with questions regarding current practices adopted in their companies.
All the interviews were conducted by at least one of the authors. The responses were documented as meeting notes, which were distributed to the interview participants.

We recognise the limitation of our case study approach, as the results of our case studies were obtained from three companies. The outcome can be specific to these companies and without further investigation, we cannot generalise the conclusion to the automotive industry.

\section{Existing architectures \label{otherworks}}

\begin{figure}[t]
\centerline{\includegraphics[width=\linewidth]{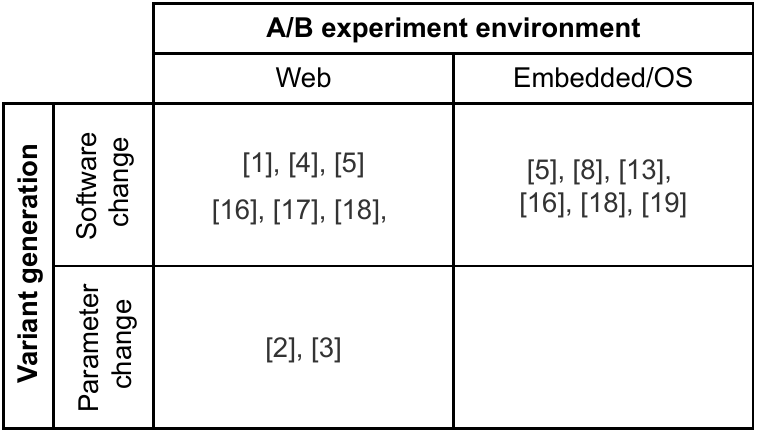}}
\caption{Existing A/B experiment framework categorised by environment and variant generation methods.}
\label{fig_litreview}
\end{figure}

In this section, we present the results from our literature review. 
We included 11 publications \cite{google2010, Eklund2012, Bosch2012, Amatriain2013, Kohavi2013, Giaimo2017, Fagerholm2017, Fabijan2018, Gupta2018, Vasthimal2019, Li2019} that focus on describing A/B experiment architectures, in both embedded software and online applications.
From our literature review, we have discovered that there is a general gap in the literature on architectures or frameworks designed specifically for automotive software.
Based on the topic, we summarise the papers into four overlapping categories. 
They are grouped firstly by their environment, i.e., embedded or web-facing. Paper \cite{Fagerholm2017, Li2019} are applicable for both groups.
We include OS and embedded applications in the same category, as they share many common challenges for instance, the devices can be offline \cite{Li2019}.
Second, we identified in these papers how a software variant is shipped to the users.
Namely, if a complete software change is required, or variants can be introduced through parameter changes.
The categories are presented in Figure \ref{fig_litreview}. As can be seen, variant introduction through parameter change is not a widely explored method within embedded software.

Although the design process is vastly different, there are a numbers of shared components for embedded and web experiment architecture. 
This includes experiment configuration, data collection, experiment analysis and metrics evaluation.
Therefore, some experiment models can be employed in various environments including web, operation systems and embedded \cite{Fagerholm2017, Gupta2018}.
Tang \etal\ \cite{google2010} and Kohavi \etal\ \cite{Kohavi2013} both report a multi layered experiment configuration system that can handle multiple A/B experiments.
Users will be assigned to A or B variant in a consistent manner \cite{Gupta2018, Vasthimal2019}.
In the web environment, this is achieved by assigning unique IDs when users visit the web pages.

Data infrastructure is a major component in any experiment framework.
All researchers include data infrastructure as part of their experiment frameworks, particularly focused on trustworthiness \cite{google2010, Kohavi2013, Fagerholm2017, Fabijan2018, Gupta2018, Vasthimal2019}. 
Such data collection is also required in embedded environments, however, is more difficult due to hardware limitations \cite{Bosch2012}. 
An experiment architecture \cite{Eklund2012} for automotive software used an on-board data storage before uploading the data through the vehicle's telemetry.
%talk about components here, focus on experiment config, data collection, data analysis.

Another key element for A/B experiments is rapid software release.
We found that all architectures for rapid experiments in an embedded environment rely on continuous deployment.
The "RIGHT Model" discuss that if a function is novel, continuous deployment might not be necessary \cite{Fagerholm2017}. 
However, most frameworks in embedded environments \cite{Eklund2012, Bosch2012, Giaimo2017} require a well-established continuous deployment process to achieve rapid experimentation loops.
Software variant release through Over-the-air(OTA) can increase delivery speed in automotive applications \cite{Eklund2012}. 
In the web environment, rapid experimentation can be achieved more flexibly through an array of mechanisms.
For example, an offline and online experiment systems in Netflix, as demonstrated by Amatriain \cite{Amatriain2013}. 
Existing data can be used to train the models before they are introduced to an online experiment, which allows faster and cheaper evaluation of software.
Another technique in increasing experiment speed is using parameter updates as mentioned by Tang \etal\ \cite{google2010}.
The A/B variants in target functions are parameterised and configured through data files.
These parameters are changed more frequently than code, which enables fast experiments provided the parameters exist.

\begin{figure*}[t]
\centerline{\includegraphics[width=\textwidth]{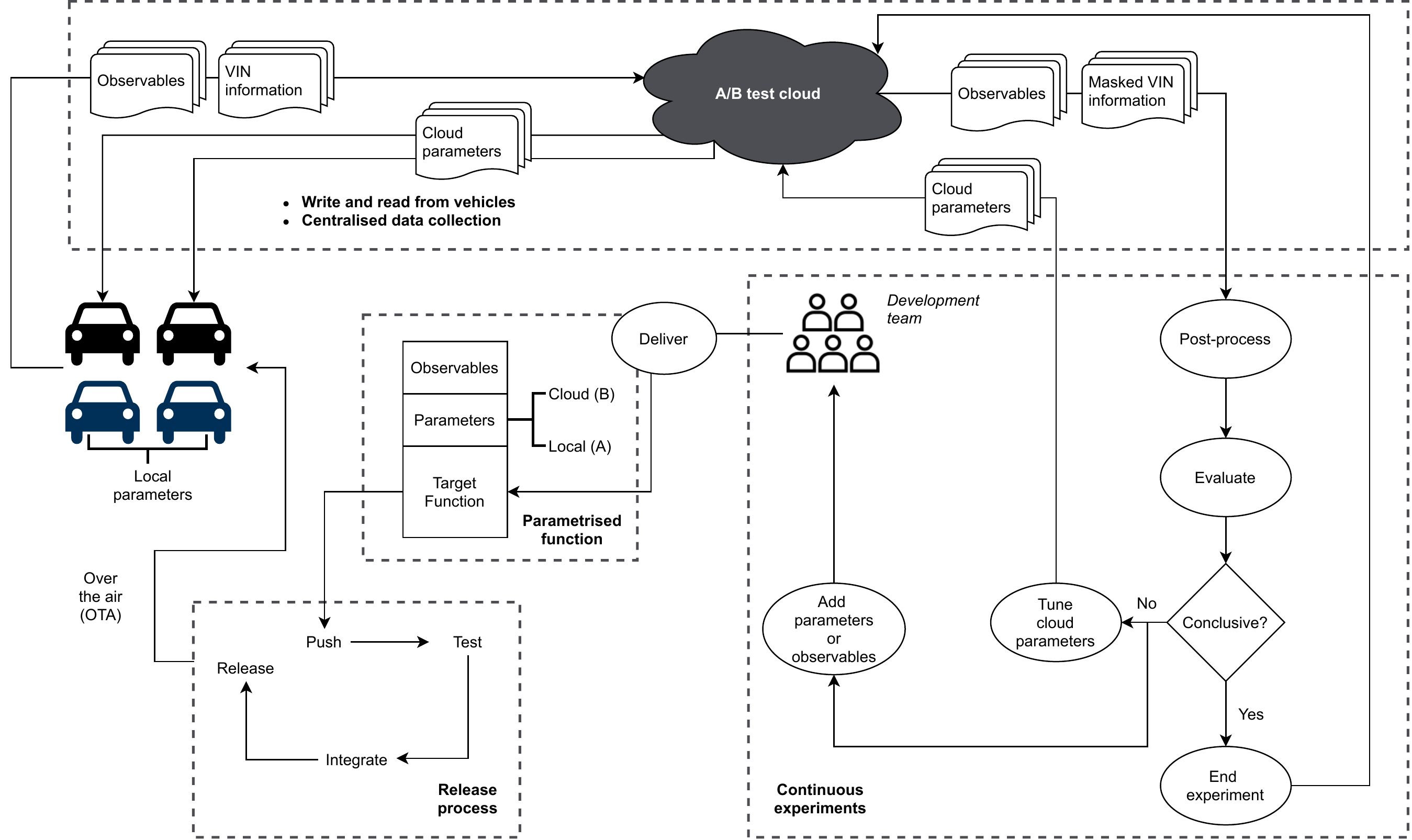}}
\caption{Process of the cloud-based A/B test architecture, illustrating the general work flow of conducting an A/B experiment with parametrised functions.}
\label{fig_process}
\end{figure*}

Furthermore, to fully utilise the benefits of A/B testing, all papers highlighted the importance of the organisational and cultural mindset of making data-driven decisions. 
Kohavi \etal\ \cite{Kohavi2013} summarise prerequisites which an organisation needs to adapt, highlighting the importance of data-driven decision making mindsets. 
In the "Experiment Growth Model" introduced by Fabijan \etal\ \cite{Fabijan2018}, all components of their A/B experimentation model become more mature as the entire organisation evolves through different stages.

\section{Architecture and case studies \label{design_main}}

%Combining understanding of A/B experimentation framework requirements and experience from automotive specific limitations, we designed a software architecture.
In this section, we present a software architecture that could enable A/B testing in the automotive domain.
%while addressing the specific challenges experienced and documented in the industry.
A hybrid architecture is presented.
%includes an on-board and a cloud component.
The essence of the architecture is to imitate an online environment for an otherwise offline application. 
In doing so, automotive A/B testing can benefit from the flexibility of online experiments We present the components of our architecture in Fig.\ref{fig_process}. 

    \subsection{System characteristics \label{desing_act}}

We present a hybrid architecture (Fig.\ref{fig_process}) combining on-board and cloud functionalities.
The system is composed of six main components. 
These are parameterised functions, a release process which most companies have in place.
There is a cloud host that writes parameters to the vehicles and collects data from the vehicles. 
Finally, a centralised data storage and pipeline for distributing the measured data.
%This A/B testing system has four processes, i.e., the software release process, selection and partition of eligible users, a cloud host which overwrites cloud parameters and the A/B experiment evaluation.

The system workflow can be described as follows. First, a function which characteristics can be defined by a list of parameters is delivered.
There are two sets of parameters embedded in the function, the local set, which is the default and the cloud set that can receive incoming values externally.
The benefit of parameterisation in A/B testing was also highlighted by \cite{google2010}.
%, as parameters can be updated more frequently than the code, and they are human readable, thus can be modified by non-experts.
A set of observables which measure function performance is also predetermined. 
The function and its parameters are delivered to a release process which will integrate with other functions and release the software to vehicles.
This release and installation of software can be done through workshop visits or OTA.
%Therefore, we do not consider OTA as an essential component of our architecture.

Once the software is introduced to the vehicles, users are identified through Vehicle Identification Numbers (VIN), which is unique and comprised of vehicle meta-data.
%the eligible users for A/B experiments will establish handshake with the cloud. 
This ensures that although the software is introduced to all cars, no experiments will be conducted unless the users are deemed eligible in advance.
%to safeguard different levels of privacy agreements.
%The eligible users can be managed over time.
%We describe the vehicle start sequence in Fig.\ref{fig_startseq}.
Upon key-on of a vehicle, a vehicle will send its VIN to the A/B test cloud.
Since the A and B groups are configured in the cloud, the test cloud will match the VIN and then return a status indicator to the vehicle.
Ineligible cars will have no match in the cloud and receive no response.
%\begin{figure}[t]
%\centerline{\includegraphics[width=\linewidth]{3_start_sequ.pdf}}
%\caption{Vehicle start sequences of A (Local) and B (Cloud) groups.}
%\label{fig_startseq}
%\end{figure}
For all eligible vehicles, they can be partitioned into A and B groups through remote configuration. 
The control group will use the functions' local parameters and the treatment group will receive cloud parameters.
Since the parameter names are predefined, the vehicle cannot accept any other values, thus increase security.
Furthermore, as the cloud parameters are blank values in the vehicles, cloud parameter change can be done remotely.
%this system is flexible in terms of further changes in the cloud parameters.
%For instance, the users can be partitioned into A/B/n groups through various subsets of cloud parameters sent remotely.
This design enables function behaviour change through parameters provided the parameters exist.
%the capability of modifying cloud parameters separately from release processes 
%In other words, the parameter change interval does not necessarily rely on a continuous deployment process which many automakers lack.
Development teams can continuously A/B test and adjust the existing parameters without complete software change and independently from the company-wide release cadence. 
A complete software update is required only when new parameters need to be added.
%If new parameters are required, it can be done through complete software updates.
%For example, one of the case study companies has a release cadence of 8 weeks through their OTA update. 
%During which, the development team could modify the cloud parameters from day to day, or week to week based on their experiment needs. 
%Thus, the architecture creates flexible continuous online experiments that are independent from the company's current release processes.

Data collection is done through the cloud and it measures a set of predefined observables.
The observables are measured and temporarily stored on board, then sent to the cloud at time intervals while driving.
This data are collected in a centralised data lake, cleaned, then distributed to development teams.
%Combining appropriate compression techniques for uploading of data to the cloud and storage of data, with appropriate determination of measurement frequency, we can scale the system relatively easily.
During a trip, time series data is collected for dynamic observables. For stationary observables, only one or a few snapshots are measured.
%\begin{figure}[t]
%\centerline{\includegraphics[width=\linewidth]{2_cadence.pdf}}
%\caption{Release cadence of a cloud-based A/B test.}
%\label{fig_time}
%\end{figure}
%The development teams can determine their analysis methods while some standardised libraries are provided.
%Standard analysis tools are important when more than one A/B experiment is ran simultaneously, since performance need to be measured in a systematised fashion.
After analysis of the A/B tests, further actions can be taken such as adjusting cloud parameters, re-partitioning A/B groups, or concluding the experiments. 
%Alternatively, parameter adjustment can be done automatically if the correlation of performance and parameter changes could be well-established or modelled.
When the experiments are concluded, the connection to the cloud will be interrupted and vehicles will invert back to the local parameters automatically.
%While receiving cloud parameters requires an active internet connection,
Moreover, the local variant always serves as a safety fallback in critical situations.

    \subsection{Case study \label{test_case}}

The first case study was performed in company A on an energy management (EM) function that was developed internally.
The function Energy Management has a local and a cloud set of parameters which determine the local and cloud energy management strategy, respectively. 
By default, the vehicle will always run the local strategy unless a connection to the cloud is established and the vehicle is eligible.
The development team delivers the software through the company's existing release process.
%These users were selected at random from corporate customers who voluntarily signed up for the experiments and their user agreement covers the data collection. 
There are 50 vehicles in this fleet of company cars driven for a total 18 month period, during which, there were 58 observables measured.
%In the fleet, there are three types of unique vehicle models.
%Internet connection through 4G were provided to the users as the function requires active internet connection to the cloud.
%Therefore, when software including EM function changes was introduced to all vehicles in the fleet, the A/B experiment was only activated for eligible users.
%Furthermore, as part of the experiment, the users were given the option to disconnect to the cloud. 
%To randomise the network and spill-over effects, the experimenters allocated A and B groups dynamically on a schedule. 
%Upon activating connections to the cloud, the local strategy was run on all vehicles for a short period to establish a baseline response.
%Then the team automated the partitioning of A/B groups through a rolling schedule while keeping the sample size of two groups balanced.
%including but not limited to net energy consumption, driving dynamics, and travel demand patterns.
The experimenters also monitor how frequently the users manually interrupt the cloud connection.

A number of automated mechanisms were put in place in the cloud to ensure data quality.
%Some observables were measured and stored as high frequency time series data while some were measured and stored as aggregate values or snapshots. 
%The experimenters were able to define the data requirements according to their experiment hypothesis.
The experimenters have access to the data collected in real-time. 
The data collected was post-processed in an automated manner in a database, while the team can also choose to export the raw data. 
The file size of data collected per week averages at 1.7 gigabytes when exported in CSV format. 
%All the analysis and evaluation were done to support decisions of further modifications to the cloud parameters when needed.
% write about how many cloud parameters there are
%B, the cloud variant of the function can be changed from the cloud without any modifications to the vehicles or the software within the vehicles.
The EM function software has dependencies on six ECUs that are mostly supplier parts.
Traditionally, changing the software means rebuilding of all these ECUs completely through suppliers and downloading the software to the vehicles physically.
The usual lead time for such changes is anywhere from three month to one year.
%With the cloud-based setup, the configuration of experiments can be done with a simple flip of a switch and the parameter changes can be done within the functional teams.
This system caters an environment where continuous experiments are independently from release processes that could be lengthy at times.
In average, the total distance travelled by all eligible users is over 18.000 kilometres per week and over 80\% of the vehicles are being driven daily.
Comparing to any test fleet, they are generating measurements at a much larger scale.

The second case study is conducted to understand the state-of-practice in company B and C.
Although neither company has experience with large scale A/B experiments, but there are commonalities in the components.
Through our interviews, it was apparent that company B and C have adopted some level of capabilities, specifically the data collection capabilities.
Company B has invested intensively in an online data collection system for their vehicles.
A set of observables are measured, their data collected and distributed to the corresponding development teams through a centralised database.
Each functional team within the company can also request for more observables to be measured from the fleet.
A similar approach was reported by company C. 
A centralised database was built to distribute high quality data in a fast manner.
The teams have the freedom to determine the sampling frequency accordingly to their measurement requirements.
%The data collected was used for various product development purposes such as remote diagnostics and predictive maintenance.

\section{Discussion \label{discussion}}

In this paper, we presented a hybrid architecture that enables continuous A/B experiments in automotive embedded software.
Comparing to the existing A/B experiment architecture, our architecture offers the flexibility of being independent from continuous deployment processes.
By allowing parameter changes, functional changes can be experimented continuously without a complete software change.

However, we foresee some potential weakness in the design and they are discussed here.
Firstly, the threshold of functional behaviour change through parameters is low comparing to a complete software change.
The system enables A/B experiments for fine tuning of functions but not complete concept changes. 
%In other words, the A/B testing can be done on relatively mature functions through our architecture.
Secondly, many parameter changes are not independent from each other in an automotive setting.
When multiple experiments are running simultaneously, the configuration of experiments becomes critical as suggested by \cite{google2010} and \cite{Kohavi2013} from their experience in online businesses.
Similar to wed-facing applications, we need to consider contradicting and hierarchical functions and their parameters.
%At the same time, some parameter configurations can be hierarchical when one function is deemed to be a sub-function of something else.
%For instance, battery cooling temperature effects, energy consumption management and climate comfort for the driver. 
%This type of parameter adjustment needs to be coordinated when both functions are being A/B tested.
Performance of contradicting or hierarchical software variants cannot be determined individually. 
Therefore, some centrally well-established and understood performance metrics need to be put in place before parallel/multiple experiments can be conducted.
Thirdly, the teams shall coordinate their experiment design when parameters or observables are shared between different functions.
Such coordination requires organisational support \cite{Fabijan2018}.
As many automotive companies are going through agile transformation \cite{Mattos2020}, the data-driven development mindsets and support structure are gradually improving.
The speed of the transformation will influence how quickly an A/B experiment framework can be implemented at scale.

Finally, receiving cloud parameters requires an active internet connection.
Although functions can be safeguarded by using local parameters as fallback, functions which require millisecond response time cannot rely on cloud connection.
%Common examples of functions which are time critical include lane keeping assist, automatic breaking and active cruise control.
A possible setup for time critical functions could be, one embeds the A and B versions of parameters in the software itself, and use the cloud to trigger the switch in between them.
As a trade-off, one will lose the freedom of tuning cloud parameters without complete software updates.

\section{Conclusion \label{conclusion}}

In recent years, some research effort was put in the adoption of A/B experiments in the automotive domain \cite{Giaimo2017, Giaimo2019, Mattos2020}. 
In this paper, we raised a research question on how to enable continuous experiments in an automotive, and presented an architecture that demonstrated such capabilities.
Through a literature review, we found that embedded experiment architectures share many components with web-facing ones, however, lack the capability of rapid changes. 
The architecture design is a hybrid A/B testing model that address many challenges in the industry. 
Comparing to existing frameworks, our hybrid architecture enable rapid software changes without compromising the high safety and security standards. 
Similar framework for automotive software A/B testing is not previously discussed in literature. 
We shared case studies of cloud-based A/B experiments at scale, which shows high potential of the parameterised hybrid architecture. 
The components of our architecture were compared with the state-of-practice of two other large automotive manufacturers. 
We found that the case study companies have applied many components, thus paving the way to an A/B experiment capable architecture.
%Furthermore, we critically discussed the pitfalls and flaws in the design. 
%In future research, we plan to continue to develop the architecture for more functions.

%\section*{Acknowledgement}

%The authors would like to thank all members of the energy management function development teams. We would also like to extend our gratitude to all interviewees at the case study companies.

%\begin{thebibliography}{00}
\bibliographystyle{IEEEtran}
\bibliography{IEEEabrv, ref.bib}

% Generated by IEEEtran.bst, version: 1.14 (2015/08/26)
\begin{thebibliography}{10}
\providecommand{\url}[1]{#1}
\csname url@samestyle\endcsname
\providecommand{\newblock}{\relax}
\providecommand{\bibinfo}[2]{#2}
\providecommand{\BIBentrySTDinterwordspacing}{\spaceskip=0pt\relax}
\providecommand{\BIBentryALTinterwordstretchfactor}{4}
\providecommand{\BIBentryALTinterwordspacing}{\spaceskip=\fontdimen2\font plus
\BIBentryALTinterwordstretchfactor\fontdimen3\font minus
  \fontdimen4\font\relax}
\providecommand{\BIBforeignlanguage}[2]{{%
\expandafter\ifx\csname l@#1\endcsname\relax
\typeout{** WARNING: IEEEtran.bst: No hyphenation pattern has been}%
\typeout{** loaded for the language `#1'. Using the pattern for}%
\typeout{** the default language instead.}%
\else
\language=\csname l@#1\endcsname
\fi
#2}}
\providecommand{\BIBdecl}{\relax}
\BIBdecl

\bibitem{Fabijan2018}
A.~Fabijan, P.~Dmitriev, C.~McFarland, L.~Vermeer, H.~H. Olsson, and J.~Bosch,
  ``Experimentation growth: Evolving trustworthy {A}/{B} testing capabilities
  in online software companies,'' \emph{Journal of Software: Evolution and
  Process}, vol.~30, no.~12, p. e2113, nov 2018.

\bibitem{google2010}
D.~Tang, A.~Agarwal, D.~O'Brien, and M.~Meyer, ``Overlapping experiment
  infrastructure: More, better, faster experimentation,'' in \emph{Proceedings
  16th Conference on Knowledge Discovery and Data Mining}, Washington, DC,
  2010, pp. 17--26.

\bibitem{Kohavi2013}
R.~Kohavi, A.~Deng, B.~Frasca, T.~Walker, Y.~Xu, and N.~Pohlmann, ``Online
  controlled experiments at large scale,'' in \emph{Proceedings of the 19th
  {ACM} {SIGKDD} international conference on Knowledge discovery and data
  mining - {KDD} {\textquotesingle}13}.\hskip 1em plus 0.5em minus 0.4em\relax
  {ACM} Press, 2013.

\bibitem{Gupta2018}
S.~Gupta, L.~Ulanova, S.~Bhardwaj, P.~Dmitriev, P.~Raff, and A.~Fabijan, ``The
  anatomy of a large-scale experimentation platform,'' in \emph{2018 {IEEE}
  International Conference on Software Architecture ({ICSA})}.\hskip 1em plus
  0.5em minus 0.4em\relax {IEEE}, apr 2018.

\bibitem{Li2019}
P.~L. Li, P.~Dmitriev, H.~M. Hu, X.~Chai, Z.~Dimov, B.~Paddock, Y.~Li,
  A.~Kirshenbaum, I.~Niculescu, and T.~Thoresen, ``Experimentation in the
  operating system: The windows experimentation platform,'' in \emph{2019
  {IEEE}/{ACM} 41st International Conference on Software Engineering: Software
  Engineering in Practice ({ICSE}-{SEIP})}.\hskip 1em plus 0.5em minus
  0.4em\relax {IEEE}, may 2019.

\bibitem{Mattos2018}
D.~I. Mattos, J.~Bosch, and H.~H. Olsson, ``Challenges and strategies for
  undertaking continuous experimentation to embedded systems: Industry and
  research perspectives,'' in \emph{Lecture Notes in Business Information
  Processing}.\hskip 1em plus 0.5em minus 0.4em\relax Springer International
  Publishing, 2018, pp. 277--292.

\bibitem{Fabijan2017a}
A.~Fabijan, P.~Dmitriev, H.~H. Olsson, and J.~Bosch, ``The benefits of
  controlled experimentation at scale,'' in \emph{2017 43rd Euromicro
  Conference on Software Engineering and Advanced Applications ({SEAA})}.\hskip
  1em plus 0.5em minus 0.4em\relax {IEEE}, aug 2017.

\bibitem{Giaimo2017}
F.~Giaimo and C.~Berger, ``Design criteria to architect continuous
  experimentation for self-driving vehicles,'' in \emph{2017 {IEEE}
  International Conference on Software Architecture ({ICSA})}.\hskip 1em plus
  0.5em minus 0.4em\relax {IEEE}, apr 2017.

\bibitem{Mattos2020}
D.~I. Mattos, J.~Bosch, H.~H. Olsson, A.~M. Korshani, and J.~Lantz,
  ``Automotive {A}/{B} testing: Challenges and lessons learned from practice,''
  in \emph{2020 46th Euromicro Conference on Software Engineering and Advanced
  Applications ({SEAA})}.\hskip 1em plus 0.5em minus 0.4em\relax {IEEE}, aug
  2020.

\bibitem{Giaimo2019}
F.~Giaimo, H.~Andrade, and C.~Berger, ``The automotive take on continuous
  experimentation: A multiple case study,'' in \emph{2019 45th Euromicro
  Conference on Software Engineering and Advanced Applications ({SEAA})}.\hskip
  1em plus 0.5em minus 0.4em\relax {IEEE}, aug 2019.

\bibitem{Dmitriev2017}
\BIBentryALTinterwordspacing
P.~Dmitriev, S.~Gupta, D.~W. Kim, and G.~Vaz, ``A dirty dozen: Twelve common
  metric interpretation pitfalls in online controlled experiments,'' in
  \emph{Proceedings of the 23rd ACM SIGKDD International Conference on
  Knowledge Discovery and Data Mining}, ser. KDD '17.\hskip 1em plus 0.5em
  minus 0.4em\relax New York, NY, USA: Association for Computing Machinery,
  2017, p. 1427–1436. [Online]. Available:
  \url{https://doi.org/10.1145/3097983.3098024}
\BIBentrySTDinterwordspacing

\bibitem{Forsberg1992}
K.~Forsberg and H.~Mooz, ``The relationship of systems engineering to the
  project cycle,'' \emph{Engineering Management Journal}, vol.~4, no.~3, pp.
  36--43, sep 1992.

\bibitem{Bosch2012}
J.~Bosch and U.~Eklund, ``Eternal embedded software: Towards innovation
  experiment systems,'' in \emph{Leveraging Applications of Formal Methods,
  Verification and Validation. Technologies for Mastering Change}.\hskip 1em
  plus 0.5em minus 0.4em\relax Springer Berlin Heidelberg, 2012, pp. 19--31.

\bibitem{Fabijan2016}
A.~Fabijan, H.~H. Olsson, and J.~Bosch, ``The lack of sharing of customer data
  in large software organizations: Challenges and implications,'' in
  \emph{Agile Processes, in Software Engineering, and Extreme
  Programming}.\hskip 1em plus 0.5em minus 0.4em\relax Springer International
  Publishing, 2016, pp. 39--52.

\bibitem{Kit04}
B.~Kitchenham, ``Procedures for performing systematic reviews,'' Department of
  Computer Science, Keele University, UK, Keele University. Technical Report
  TR/SE-0401, 2004.

\bibitem{Eklund2012}
U.~Eklund and J.~Bosch, ``Architecture for large-scale innovation experiment
  systems,'' in \emph{2012 Joint Working {IEEE}/{IFIP} Conference on Software
  Architecture and European Conference on Software Architecture}.\hskip 1em
  plus 0.5em minus 0.4em\relax {IEEE}, aug 2012.

\bibitem{Amatriain2013}
X.~Amatriain, ``Beyond data: from user information to business value through
  personalized recommendations and consumer science,'' \emph{Proceedings of the
  22nd ACM international conference on Information \& Knowledge Management},
  2013.

\bibitem{Fagerholm2017}
F.~Fagerholm, A.~S. Guinea, H.~M{\"a}enp{\"a}{\"a}, and J.~M{\"u}nch, ``The
  {RIGHT} model for continuous experimentation,'' \emph{Journal of Systems and
  Software}, vol. 123, pp. 292--305, jan 2017.

\bibitem{Vasthimal2019}
D.~K. Vasthimal, P.~K. Srirama, and A.~K. Akkinapalli, ``Scalable data
  reporting platform for a/b tests,'' in \emph{2019 {IEEE} 5th Intl Conference
  on Big Data Security on Cloud ({BigDataSecurity}), {IEEE} Intl Conference on
  High Performance and Smart Computing, ({HPSC}) and {IEEE} Intl Conference on
  Intelligent Data and Security ({IDS})}.\hskip 1em plus 0.5em minus
  0.4em\relax {IEEE}, may 2019.

\bibitem{Runeson2008}
P.~Runeson and M.~Höst, ``Guidelines for conducting and reporting case study
  research in software engineering,'' \emph{Empirical Software Engineering},
  vol.~14, no.~2, pp. 131--164, dec 2008.

\end{thebibliography}

%\end{thebibliography}

\end{document}